\documentclass[12pt,twoside,aps,preprint,preprintnumbers,showpacs,showkeys,amsmath,amssymb]{revtex4}
\usepackage{amsthm,amscd}
\usepackage[all]{xy}
    \theoremstyle{definition}
\newtheorem{groupoid}{Definition}[section]

    \theoremstyle{plain}

\newtheorem{olem}[groupoid]{Lemma}

\swapnumbers
    \theoremstyle{definition}
\newtheorem{isometry}{Definition}[section]
\newtheorem{ddef}[isometry]{Definition}
\newtheorem{dnote}[isometry]{Noteworthy}

\newtheorem{dex}[isometry]{Example}
\newtheorem{dassu}[isometry]{Assumption}
        \theoremstyle{plain}
\newtheorem{dthm}[isometry]{Theorem}
\newtheorem{dlem}[isometry]{Lemma}

\numberwithin{equation}{section}

\newcommand{\C}{\mathbb{C}}\newcommand{\Real}{\mathbb{R}}
\newcommand{\N}{\mathbb{N}}

\newcommand{\ie}{\textit{i.e.}\,}

\newcommand{\obj}{\operatorname{obj}}

\newcommand{\trace}{\operatorname{tr}}

\newcommand{\im}{\operatorname{im}}
\newcommand{\End}{\operatorname{End\,}}
\newcommand{\id}{\operatorname{id}}

\newcommand{\der}{\operatorname{der}}

\newcommand{\half}{\textstyle{\frac{1}{2}}}

\newcommand{\V}{\mathbb{V}}

\newcommand{\Obs}{\operatorname{Obs}}
\newcommand{\Cl}{\mathcal{C}\ell}

\newcommand{\vel}{\mathbf{v}}\newcommand{\cel}{\mathbf{c}}
\newcommand{\uel}{\mathbf{u}}\newcommand{\xel}{\mathbf{x}}
\newcommand{\wel}{\mathbf{w}}
\newcommand{\zel}{\mathbf{0}}\newcommand{\ael}{\mathbf{a}}

\newcommand{\olie}{\mathfrak{o}}
\newcommand{\bt}{\begin{tabular}{c}}
\newcommand{\et}{\end{tabular}}
\def\bf{\begin{flushright}}
\def\ef{\end{flushright}}

\begin{document}\preprint{Fifth Workshop Applied Category Theory,
Graph-Operad-Logic, Merida May 2006}
\title{THE LORENTZ BOOST-LINK IS NOT UNIQUE.\\
Relative velocity as a morphism\\in a connected groupoid category of
null objects}
\author{\textbf{Zbigniew Oziewicz}\footnote{Supported by el Consejo Nacional de Ciencia y
Tecnolog\'{\i}a (CONACyT de M\'exico), Grant \# U 41214 F. A member
of Sistema Nacional de Investigadores in M\'exico, Expediente \#
15337.}}\affiliation{\emph{Universidad Nacional Aut\'onoma de
M\'exico\\Facultad de Estudios Superiores Cuautitl\'an\\Apartado
Postal \# 25\\C.P. 54714 Cuautitl\'an Izcalli\\Estado de M\'exico,
M\'exico}}\email{oziewicz@servidor.unam.mx}
\date{Received: August 7, 2006}
\begin{abstract}
The isometry-link problem is to determine all isometry
transformations among given pair of vectors with the condition that
if these initial and final vectors coincide, the transformation-link
must be identity on entire vector space. Such transformations-links
are said to be the pure, or the boost, transformations. The main
question, possed by van Wyk in 1986, is: how many there are
pure-isometry-links among given pair of vectors of the same
magnitude?

In the first part of this essay we provide the complete solution for
the link problem for arbitrary isometry, for any dimension and
arbitrary signature of the invertible metric tensor. We are proving
that in generic case each solution of the link problem is not given
uniquely by the given initial and final vectors; each solution needs
the third vector called the privileged or preferred vector: the
triple of vectors determine the unique pure isometry. If triple of
vectors is coplanar the isometry-link is given uniquely by initial
and final vectors. Non-planar systems gives, in general, infinite
set of isometry-link solutions.

We apply these considerations for pure Lorentz transformations, for
the Lorentz boost, parameterized by relative velocity, and we are
showing that the isometric pure Lorentz transformation-link is not
given uniquely by the initial and final vectors. Lorentz's boost
needs a choice of the preferred time-like observer. This leads to
non-uniqueness of the relative velocity among two reference systems,
that was apparently not intention of Einstein in 1905.

In order to have axiomatically the unique relative velocity among
pair of massive bodies, we propose a connected groupoid category of
massive bodies in mutual motions. Groupoid is a small category in
which all morphisms are isomorphisms. In groupoid category of
massive bodies, each morphism is the unique relative velocity (not
non-unique isometric Lorentz transformation). We propose to consider
the relative velocity as the primary concept with dichotomy: the
unique velocity-morphism, \textit{versus} the morphisms given by a
set of non-unique isometric Lorentz boost-links.

Presently, as during the XX century, the Lorentz covariance is the
cornerstone of physical theory. Our main conclusion is:
observer-dependence (and -independence), and the
Lo\-rentz\--co\-var\-iance (and invariance), are different concepts.
The same statement holds in physics with absolute simultaneity:
XVII-century observer-independence is not the same as the XX-century
Galilean-group-invariance-covariance. We think that Lorentz
covariance will dwindle in importance.

The kinematics of groupoid-category-relativity is ruled by Frobenius
algebra, whereas the dynamics needs the Fr\"olicher-Richardson
non-associative algebra.\end{abstract}
\maketitle


\pagestyle{myheadings}\markboth{\quad\hrulefill\quad Zbigniew 
Oziewicz\quad}{\quad Relative velocity as
morphism\quad\hrulefill\quad}

\hyphenation{idem-po-tent a-do-p-ted E-vi-den-tly}

\section{\label{} Relativity concepts \lowercase{in terms of} category theory}
\begin{quotation} Science is the meaning of reality, not the photograph
of reality. \bf H. Bergman 1929, cited in \textit{Boston Logical
Studies} [1974, p. 400].\ef\end{quotation}


Relativity of the space, \ie many relative spaces of Galileo
[Galileo 1632], and relativity of time, \ie many relative
simultaneity of Albert Einstein [Einstein 1905], and many relative
proper-times of Minkowski [Minkowski 1908], all were formulated long
before the first concepts of the category theory were invented in
1945 [Eilenberg and Mac Lane 1945].

The relativity principle was known since Kopernicus (Copernicus)
[1543] and Galileo [1632]. Poincar\'e [1904], Lorentz [1904] and
Einstein [1905], formulated the relativity principle in the
following form: laws of physics, including electromagnetic
phenomena, should be independent of the choice of the inertial
system of reference, should be inertial-observer-free. This
principle is said to be the principle of the special (inertial)
relativity, with constant relative velocities. Einstein extended
this to the principle of general relativity, stating that laws of
physics, including gravity, should be observer-free relative to
\textit{all} reference systems, including all non-inertial systems,
accelerated, deformed, rotated.

Brillouin stressed in [1970] that the reference system must be
physical \textit{massive} body, and not mathematical coordinate
system. The \textit{massless} cosmic microwave background radiation,
often considered as the `natural preferred frame of reference', see
e.g. [Braxmaier et al. 2002], can not be considered as the physical
observer.

The crucial is the mathematical model of the physical massive body.

The physical massive body, with mass distribution, is considered to
be located in the primitive absolute configuration space. Lawvere
[2002], in the science of materials, consider the topos category of
spaces, with relations between body and absolute space.

Wagh [2006] propose universal relativity that is going beyond the
gravity, and for all reference systems, in framework of point-free
topology. It is proposed that a physical reference system is given
by a complete lattice, called \textit{frame}, and it is noted that
category of such physical bodies/frames is not a topos category.

The above attempts of describing the physical massive bodies in a
topos category, or in non-topos category, takes as the primitive
concepts the space, location, topology, distance. For another
approach see also [Laudal 2005].

Contrary to the above philosophy, here we propose to consider the
relative velocity as the most primitive concept. The relative
velocity, not necessarily constant, being the categorical morphism
in a connected groupoid category of massive physical bodies.
Groupoid is a small category in which all morphisms (in our case all
relative velocity-morphisms) are isomorphisms. In this philosophy,
the relative spaces and relative (proper)-times are derived
secondary concepts.

\subsection{Overlook}
In the Einstein special relativity theory an object is an observer
that is understand just as a reference frame (a basis) or as the
coordinate-system. A morphism among such objects is the isometric
inhomogeneous Lorentz coordinate transformation. A relative velocity
is reciprocal (skew-symmetric), and is parameterizing the isometric
Lorentz-boost among two bodies. In this essay we are going to show
that such 'isometric' relative velocity is
preferred-laboratory-dependent, see
\eqref{ternary}-\eqref{ternary0}, and the addition of such
\textit{ternary} velocities must be non-associative.

In the Lorentz-group-free categorical relativity theory, proposed
here, an observer is (1+3)-split-idempotent, and a morphism between
observers is an intrinsic unique binary relative velocity-morphism,
that is not skew-symmetric (not reciprocal). This imply that the
addition of binary velocities must be associative, but a change of
an observer-idempotent is not an isometry (not Lorentz
transformation).

Albert Einstein in 1905 made reciprocity assumption that the inverse
of the addition of relative velocities must be the same as for the
Galilean absolute simultaneity. We are going to show in this essay
that the reciprocity axiom, that the mutual speeds of two reference
systems differ by sign solely, leads to ternary relative velocity
(with respect to the preferred time-like (massive) reference
system), with non-associative addition. We think that the
reciprocity axiom belongs to absolute-simultaneity physics and is
not needed by relativity theory.

The central notion of our new proposal is an enriched groupoid
category of most general non-inertial observers (time-like vector
fields), and an operator algebra generated by
$(1+3)$-splits-idempotents, that hopefully is a Frobenius algebra.
Thus the primary notion is not a space-time, but an operator
algebra, in some analogy to formulation of quantum mechanics outside
of the Hilbert-space model. This is background independent
formulation, without a fixed background spacetime. A spacetime is
seen as a modul/comodule over this primary operator algebra.

Starting from the associative addition of binary relative
velocities-morphisms, and selecting the preferred massive reference
system, one can derive the non-associative addition of ternary
Lorentz-boost-velocities.This derivation will be published
elsewhere.

The different relative velocities (the unique binary
velocity-morphisms in a group\-o\-id category of observers, versus
many ternary Lorentz-boost-links) and different additions (the
Lo\-rentz\--group-free associative addition, \textit{versus}
Lo\-rentz-boost-based non-associative addition) are related by a map
(in Frobenius algebra) that do not extend to the Lorentz
transformation.

The main conclusion: the observer-independence and the
Lorentz-group-invariance must not be identified as was believed
during XX century.

\begin{acknowledgments} Extensive discussions and inspiring correspondence
over last two years, 2004-2006, with William Page (Kingston, Canada)
are most gratefully acknowledged.

The present essay is inspired also by hot and long email-discussion
we have had in 2004 with David Finkelstein (Georgia Institute of
Technology) and with Abraham Ungar (North Dakota State University),
on uniqueness of the Lorentz boost, and on the meaning of the
Lorentz relativity transformation. This subject is presented as
coordinate-dependent and basis-dependent matrix calculus in Ungar's
monograph [Ungar 2001]. Instead I was attempting to be basis-free
and coordinate-free. I am thankful to Abraham Ungar and to David
Finkelstein for this hot email correspondence.

I like to thank Bernard Jancewicz (Uniwersytet Wroc{\l}awski,
Poland), and to Professor Sanjay M. Wagh (Central India Research
Institute, Nagpur), for inspiring correspondence and for pointing
important references.\end{acknowledgments}

\section{Isometry is coordinate-free and basis free}\label{seciso}
The Lorentz relativity transformation is an \textit{isometry}. In
the last Sections of this essay we propose an alternative
mathematical relativity theory that does not need the Lorentz
isometric transformations. The alternative formulation we call the
Lorentz-group-free relativity. It is the groupoid category of
reference systems, where morphisms are not isometries. In order to
compare these two mathematical formulations of relativity theory,
one with the group of Lorentz relativity transformations, and
another without of the group of the Lorentz relativity
transformations, we wish to clarify what is the coordinate-free and
basis-free isometry. Now, in XXI century, the majority of the
physics community are not familiar with the coordinate-free
discussion. A tensor as a coordinate-free concept is alien, and
coordinate-free reasoning is considered as the defect `enveloping
the physical ideas'. Nowadays, year 2006, the physics community
demand the presentation explicitly in `sacred' coordinates and
basis-dependent matrices of XVIII and XIX centuries. Coordinates and
matrices (matrix of the boost, Dirac matrix, etc) is considered as
the only legitimate language of physics, maybe the language for all
science? Contrary to this mainstream adoration of coordinates and
bases (= frames), here we wish explain that the concept of an
isometry is coordinate-free and basis-free. We believe that
coordinates and/or bases introduce something irrelevant into
scientific thinking, and obscure both, the mathematical and the
physical ideas. For discussion of the obscure meaning of the
coordinate systems we refer to [Hermann Weyl 1921, 1926].

Each Lorentz and Poincar\'e (inhomogeneous Lorentz) transformation
is an isometry. The Lorentz and Poincar\'e groups are the groups of
symmetries of the metric-\textit{tensor} of the
\textit{empty}-space-time, no matter. The question is: why
relativity transformation from one \textit{massive} reference system
to another massive reference system must be obligatory the isometry
of the \textit{empty}-energy-less spacetime?

In order to understand the physical meaning of the Lorentz
relativity transformation (also viewed as coordinate-free), and
therefore the physical meaning and the interpretation of the
Einstein's special relativity theory, it is desirable to understand
the mathematical concept of isometry outside of the relativity
theory.

So, what is an isometry? Consider a module over associative
commutative algebra, or a vector space $V$ over a field, of
arbitrary finite dimensionality that is irrelevant for the
discussion in this Section. Consider for a simplicity a vector
space, $\Real$-space $V,$ no relativity theory yet. Let $V^*$ be a
dual $\Real$-space, and let $g\in V^*\otimes_\Real V^*,$ be
symmetric metric tensor, considered as the map $V\otimes
V\rightarrow\Real,$ and as the isomorphism, $g=g^*:V\rightarrow
V^*.$ The signature of $g$ is arbitrary and irrelevant for the
discussion in the present Section. Let $A,B\in V$ be vectors, then
the scalar product is usually denoted by $A\cdot B=g(A,B)\in\Real.$
However, in fact this means $g(A\otimes B),$ because the domain of
metric \textit{tensor} $g$ are second rank tensors, like $A\otimes
B,$ and not Cartesian pairs $(A,B)\in V\times V.$ We will use
$A\cdot B$ for brevity. The scalar product $g$ of vectors,
$g:V\rightarrow V^*,$ extends by algebra homomorphism to scalar
product $g^\otimes$ of all tensors, $g^\otimes:V^\otimes\rightarrow
V^{*\otimes},$ and extends to scalar product $g^\wedge$ of all
Grassmann multi-vectors, $g^\wedge:V^\wedge\rightarrow V^{*\wedge}.$

\begin{dex} Let $A\wedge B$ and $P\wedge Q$ be simple bi-vectors
in $V^{\wedge 2}\equiv V\wedge V.$ Then
\begin{gather*}(A\wedge B)\cdot(P\wedge Q)\equiv
g^\wedge((A\wedge B)\otimes(P\wedge Q))=(A\cdot P)(B\cdot Q)-(A\cdot
Q)(B\cdot P).\end{gather*}\end{dex}

\begin{isometry}[Isometry]\label{isom0} An endomorphism $L$ of $\Real$-space $V,$
$L\in\End\,V=V\otimes V^*,$ is said to be $g$-isometry if the scalar
product is invariant,
\begin{gather}\forall\;A,B\in V,\qquad g((LA)\otimes(LB))=(LA)\cdot(LB)=
A\cdot B,\label{isom2}\\L^*\circ g\circ
L=g.\label{isom3}\end{gather} The set of all $g$-isometries is a Lie
group, denoted by $O_g,$
\begin{gather}\dim\,V=n\quad\Longrightarrow\quad\dim(O_g)=\dim(V^{\wedge 2})
=\binom{n}{2}=\frac{n(n-1)}{2}.\end{gather}\end{isometry}

If $P\in V$ is a vector (basis-free!), and $\alpha\in V^*$ is a
covector, then $P\otimes\alpha\in\End\,V,$ is said to be simple
endomorphism, with quadratic minimal polynomial (kind of
idempotency),
\begin{gather}\trace(P\otimes\alpha)=\alpha P,\quad(P\otimes\alpha)^2=
(\alpha P)(P\otimes\alpha),\\
g\circ(P\otimes\alpha)=\{(gP)\otimes(g^{-1}\alpha)\}\circ g.
\end{gather}

In this essay we are interested in the parametrization of isometry
in terms of basis-free vectors, and in terms basis-free Grassmann
bivectors, keeping in mind the application for the parametrization
of the Lorentz boost in terms of `the vector of relative velocity'
that will be presented in the basis-free way in the next Sections.

One can try parameterize isometry $L\in O_g$ in terms of the single
vector $P$ as, \begin{gather}L_P=\id-P\otimes\alpha,\end{gather} for
unknown covector $\alpha.$ Inserting this expression into the
definition of the isometry \eqref{isom3}, we get the unique solution
for covector $\alpha,$ provided that $P^2\neq 0.$ Such
single-vector-parameterized-isometry appears to be the
ray-parameterizations that depends on one-dimensional subspace span
by non light-like vector field (or a vector) $P.$ Moreover, such
isometry must be unipotent = reflection,
\begin{gather}L_P=\id-2\frac{P\otimes gP}{P^2}\qquad\Longrightarrow\qquad
(L_P)^{-1}=L_P.\label{refl1}\end{gather}

The next step is to look for the parametrization of isometry in
terms of the pair of vectors, the triple of vectors \ldots, etc, \ie
to consider the sum of the simple endomorphism
$L=\id-\sum_i(P_i\otimes\alpha^i).$ The sum of simple endomorphism
(simple tensors) that is not simple is said to be an entanglement.
For more details about entanglement, see the Lecture by Guillermo
Morales-Luna in this Volume.

Consider the parametrization of isometry in terms of the sum of two
idempotents. This is the following ansatz for two unknown covectors,
$\alpha$ and $\beta,$ or for four unknown scalars, $a,b,c,e,$ as was
considered by van Wyk in 1958,
\begin{gather}L=\id-P\otimes\alpha-Q\otimes\beta,\qquad P\wedge Q\neq
0,\quad\alpha\wedge\beta\neq 0,\notag\\
g^{-1}\alpha=aP+bQ,\quad g^{-1}\beta=cP+eQ.\label{s5}\end{gather}
Inserting above ansatz into \eqref{isom2}-\eqref{isom3} we get the
following system of three scalar equations
\begin{gather}(aP+cQ)^2=2a,\quad(bP+eQ)^2=2e\quad
\left((a+b)P+(c+e)Q\right)^2=2(a+b+c+e).\label{s1}\end{gather} One
can solve above system \eqref{s1} in terms of the next ansatz in
terms of one unknown $P$- and $Q$-dependent scalar field
$\gamma\equiv\gamma_{P,Q},$
\begin{gather}a=\frac{Q^2}{\gamma+1},\quad b=-1-\frac{P\cdot Q}{\gamma+1},
\quad e=\frac{P^2}{\gamma+1},\quad c=+1-\frac{P\cdot
Q}{\gamma+1}.\label{s2}\end{gather} Then, the isometry condition
\eqref{isom2}-\eqref{isom3} and the system of equations
\eqref{s1}-\eqref{s2} is reduced to the following single equation
for a scalar $\gamma_{P\wedge Q},$
\begin{gather}(\gamma_{P\wedge Q})^2=1-(P\wedge Q)^2,\qquad(P\wedge Q)^2\leq 1.
\label{s3}\end{gather} The last condition on the magnitude of the
bivector $P\wedge Q$ assure that the scalar field $\gamma_{P\wedge
Q}$ is real. Isometry parameterized by the pair of vectors,
$\{P,Q\},$ \eqref{s5}-\eqref{s3} possess the third order minimal
polynomial (distinct from reflection \eqref{refl1}), and this
minimal polynomial depends, through \eqref{s3}, on the magnitude of
the bivector $(P\wedge Q)^2$ only,
\begin{gather}(L-1)\left\{L^2+2(\gamma-2)(L+1)\right\}=0.\label{mp}
\end{gather}

The parametrization \eqref{s5} in terms of the sum of two simple
endomorphisms, is in fact the simple-bi-vector parametrization
\begin{gather}V^{\wedge 2}\ni P\wedge Q\quad\hookrightarrow\quad
L_{P\wedge Q}\in O_g.\end{gather} This can be seen as follows. The
vector space of Grassmann bi-vectors $V^{\wedge 2},$ inside of the
Clifford algebra, $V^{\wedge 2} \subset\Cl(V,g),$ is the Lie algebra
of the Lie group $O_g,$ $\dim(O_g)=\dim(V^{\wedge 2}).$ There is the
Lie algebra morphism $M: V\wedge V\longrightarrow V\otimes V^*,$
that we are going to describe now. For each vector $v\in V,$ $gv\in
V^*,$ we denote by $i_{gv}\in\der(\text{Grass})$ the graded
derivation of the Grassmann algebra $V^\wedge.$ For each bivector
$b\in V^{\wedge 2}$ (not necessarily simple), and for each vector
$v\in V,$ the linear map $M$ is defined as follows
\begin{gather}V^{\wedge 2}\ni b\quad\longrightarrow\quad M_b\,v\equiv-i_{gv}\,b
\quad\in V.\end{gather} One can show that the image of the above
map, $\im M\subset\End V,$ are trace-less and $g$-skew-symmetric
endomorphisms. Therefore $M$ extends to the Lie algebra isomorphism
from the Lie algebra of bivectors $\olie_g,$ to the Lie algebra of
trace-less $g$-skew-symmetric endomorphism of $V.$

This means that an isometry $L\in O_g$ can be generated by
bi-vectors,
\begin{gather}M_{P\wedge Q}\equiv P\otimes gQ-Q\otimes gP\quad\in\End\,V.
\end{gather}
The endomorphism $M_{P\wedge Q}$ possess the third order minimal
polynomial, and therefore there is no surprise that the following
binomial coincide with the parametrization \eqref{s5}-\eqref{s3},
\begin{gather}L_{P\wedge Q}=\id+M_{P\wedge Q}+\frac{1}{\gamma+1}(M_{P\wedge Q})^2.
\label{bino}\end{gather}

In this way we proved the following theorem, that holds for
$2\leq\dim\,V,$ and for arbitrary signature of the metric tensor
$g.$
\begin{dthm}Let the simple bivector $P\wedge Q$ satisfy the following
condition \eqref{condi}, and the scalar $\lambda$ is given as
follows
\begin{gather}(P\wedge Q)^2\leq 1,\qquad\gamma_{P\wedge Q}\equiv\sqrt{1-(P\wedge Q)^2}.
\label{condi}\end{gather}

Let an endomorphism $L_{P\wedge Q}\in\End V,$ be defined as follows
\begin{align}L_{P\wedge Q}&=\id+P\otimes gQ-Q\otimes gP+\frac{1}{\gamma+1}
(P\otimes gQ-Q\otimes gP)^2\notag\\
&=\id-\frac{1}{\gamma+1}P\otimes\left\{Q^2gP- (P\cdot
Q+\gamma+1)gQ\right\}-\frac{1}{\gamma+1}Q\otimes\left\{P^2gQ-(P\cdot
Q-\gamma+1)gP\right\}.\label{isom}\end{align} Then, the above
endomorphism \eqref{isom} is the $g$-isometry, $L_{P\wedge Q}\in
O_g,$ with the minimal polynomial \eqref{mp}. An inverse is
$\{L_{P\wedge Q}\}^{-1}=L_{Q\wedge P}.$\end{dthm}
\begin{proof} It is sufficient to verify the following equalities for all
vectors $K,N,$
\begin{gather}L_{P\wedge Q}\circ L_{Q\wedge P}=\id,\qquad
(L_{P\wedge Q}K)\cdot(L_{P\wedge Q}N)=K\cdot N.\end{gather} One must
use condition \eqref{condi}.\end{proof}

The discussion which follows in the next Sections will be based on
the technical knowledge of the above coordinate-free and basis-free
simple-bivector-parametrization of $g$-isometry \eqref{isom} with
condition \eqref{condi}. The isometry $L_{P\wedge Q}\in O_g$
\eqref{isom} is a second order polynomial in generating bivector
from the Lie algebra $P\wedge Q\in{\olie}_g,$ \eqref{bino}. In
explicit expression for the isometry $L_{P\wedge Q},$ \eqref{isom},
we see the vectors $P$ and $Q$ as separated, and the bivector
$P\wedge Q,$ as the \textit{only} intrinsic variable, see
\eqref{bino}, is hidden.

The $g$-isometry \eqref{isom}, imply in particular that,
\begin{align}L_{P\wedge Q}P&=\left\{1+P\cdot Q-\frac{1}{\gamma+1}(P\wedge Q)^2
\right\}P-P^2Q,\\
L_{P\wedge Q}Q&=\left\{1-P\cdot Q-\frac{1}{\gamma+1}(P\wedge Q)^2
\right\}Q+Q^2P.\end{align}

Let $P^2Q^2=+1.$ Then in \eqref{condi}, $\gamma=|P\cdot Q|,$
\begin{gather}L_{P\wedge Q}P=\begin{cases}2(P\cdot Q)P-P^2Q&\quad\text{if}
\quad 0\leq P\cdot Q,\\\qquad-P^2Q&\quad\text{if}\quad P\cdot Q\leq
0\end{cases}\\
L_{P\wedge Q}Q=\begin{cases}\quad\qquad Q^2P&\quad\text{if}\quad
0\leq P\cdot Q,\\-2(P\cdot Q)Q+Q^2P&\quad\text{if}\quad P\cdot Q\leq
0\end{cases}\end{gather}

Ivanitskaja consider the Lorentz group $O(1,3),$ and ansatz in terms
of not simple bivectors, similar to,
$L=\id-\sum_{i=1}^{i=4}(P_i\otimes\alpha^i),$ [Ivanitskaja 1979,
Chapter VI, page 292, formula (26.15)].

\subsection{Bivector has many presentations}\label{V-presen}
Each simple Grassmann bivector is $SL2$-invariant, \ie has many
$SL2$-presentations in terms of vectors. For scalars $a,b,c,e,$
$\bigl(\begin{smallmatrix}a&b\\c&e\end{smallmatrix}\bigr)\in SL2,$
we have,
\begin{gather}ae-bc=1\quad\Longleftrightarrow\quad(aP+bQ)\wedge(cP+eQ)=
P\wedge Q.\label{presen}\end{gather} We need the following
idempotent for non light-like vector $P^2\neq 0,$
\begin{gather}p\equiv\frac{P\otimes gP}{P^2}\quad\Longrightarrow\quad
p^2=p,\quad\trace p=1.\label{idem}\end{gather} The above freedom of
the presentation \eqref{presen}, can be used to fulfill the
orthogonality of vectors,
\begin{gather}P\wedge Q=P\wedge(Q-fP),\quad
P\cdot(Q-fP)=0\quad\Longrightarrow\quad f=\frac{P\cdot Q}{P^2},\\
W\equiv(\id-p)\,Q=Q-\frac{P\cdot Q}{P^2}P,\quad\text{is orthogonal
to $P,$}\quad W\cdot P=0.\end{gather}

In what follows it is convenient to use different presentations of
the bivectors,\begin{gather}P\wedge Q=P\wedge(\id-p)\,Q=P\wedge
W,\quad\text{where}\quad P^2\neq 0\quad\text{and}\quad P\cdot
W=0.\end{gather}

We stress that the change of the presentation in terms of different
vectors does not change the bivector in question, and does not
change the isometry, $L_{P\wedge Q}=L_{P\wedge W}.$ This is the
change of the parametrization of the same isometry. In the
orthogonal presentation, $P\cdot W=0,$ of the simple bivector
$P\wedge W,$ we have
\begin{gather}(P\wedge W)^2=P^2W^2\leq 1,\quad\gamma=\sqrt{1-P^2W^2},\notag\\
L_{P\wedge W}=\id+P\otimes gW-W\otimes gP-\frac{1}{\gamma+1}
\left(W^2P\otimes gP+P^2W\otimes gW\right),\notag\\
L_{P\wedge W}P=-P^2W+\left(1-\frac{P^2W^2}{\gamma+1}\right)P,\notag\\
L_{P\wedge W}W=
+W^2P+\left(1-\frac{P^2W^2}{\gamma+1}\right)W.\label{W}\end{gather}
The bivector is presentation-independent, therefore the magnitude of
the bivector is presentation-independent also.

\section{Isometry as a link: the complete solution} In this
Section we consider the link equation for the isometry. This is a
mathematical problem outside of relativity theory, and is formulated
for dimension $\geq 2,$ and for arbitrary signature of the
invertible metric tensor field $g.$

There is an obvious fact in group theory, emphasized by Wigner
[1939], that every group transformation from an initial
source-vector $R,$ to a target-vector $S,$ is up to the Wigner
little subgroups of $R$ and $S,$ known also as stabilizers, the
stability subgroups. For example, the Wigner little subgroup of the
isometry group $O_g$ of a vector $R$ is defined as follows
\begin{gather}O^R\equiv\{k\in O_g,\;kR=R\}.\end{gather}
Therefore the link equation $LR=S$ for the given vectors $R$ and
$S,$ such that \begin{gather}(R-S)\cdot(R+S)=R^2-S^2=0,\end{gather}
and for unknown isometry $L\in O_g,$ has the little-group-ambiguity,
\begin{gather}\forall\,r\in O^R\quad\text{and}\quad\forall\,s\in O^S,\quad
LR=(s\circ L\circ r)R=S.\end{gather} If an isometry $L=L(R,S)$ solve
the link equation $LR=S$, then also $(s\circ L\circ r)$ is a
solution, and there is infinite set of links up to the
Wigner-stabilizer ambiguity. Such Wigner-little-group non-uniqueness
means that the solution of link-equation $LR=R$ need not to be an
identity for $R=S,$ $L(R,R)\neq\id\in\End V.$ In order to be
stabilizer-independent we restrict the solutions of the link
equation $LR=S,$ by pure-link condition, $L(R,R)=\id,$
$L|(R=S)=\id.$

\begin{ddef}[Link equation and boost] Let $R$ and $S$ be given vectors
such that $R^2=S^2,$ \ie $(R-S)\cdot(R+S)=0.$ The link equation for
the unknown $g$-isometry $L\in O_g,$ is $LR=S.$ The isometry-link
solution $L=L(R,S),$ such that, $L(R,R)=\id,$ is said to be a
pure-isometry-link, or a boost.\end{ddef}

The question is about the uniqueness of the isometry-link: how many
there are the different pure-isometry-links, $L=L(R,S)\in O_g,$ for
the given vectors $R$ and $S,$ such that $(R-S)\cdot(R+S)=R^2-S^2=0$
?

When an isometry $L\in O_g$ is a reflection, $L^2=\id,$
parameterized by a single unknown non-zero vector $P,$ the link
equation $L_PR=S,$ for given conditions, $(R-S)\cdot(R+S)=0$ and
$(R-S)^2\neq 0,$ has solution up to non-zero scalar $\mu\neq 0,$
$P=\mu(R-S).$ This `Galilean subtraction', $R-S,$ can be interpreted
as a `relative parametria from $R$ to $S$\,',
\begin{gather}L_PR=S\quad\Longleftrightarrow\quad P=\mu(R-S).\label{refl2}
\end{gather}

Our problem in this Section is to find the most general solution for
the link equation for $g$-isometry parameterized by simple-bivector
\eqref{isom}, $L_{P\wedge Q}R=S.$ This is a problem to find
\textit{all} simple bivectors $P\wedge Q,$ for the given pair of
vectors $R$ and $S,$ subject to constraint $(R-S)(R+S)=0.$ This
problem has the separate complete solution for the generic case when
$(R-S)^2\neq 0,$ and another solution for singular case $(R-S)^2=0.$

\begin{dassu}\label{assum} In what follows we assume that the system of
three non-zero vectors $\{R,S,P\},$ is subject to the following
conditions,
\begin{gather}(R-S)\cdot(R+S)=0,\quad\text{and}\quad(R-S)^2\neq 0,\notag\\
P\cdot(R+S)\neq 0,\quad\text{and}\quad P\wedge(R-S)\neq 0,\notag\\
\{P\wedge(R-S)\}^2+\{P\cdot(R+S)\}^2=P^2(R-S)^2+4(P\cdot R) (P\cdot
S)\neq 0.\label{condition}\end{gather} The non-zero scalar
$\mu=\mu(P,R\pm S),$ for the above system of three vectors is
defined as follows
\begin{gather}\mu\equiv\frac{2P\cdot(R+S)}{\{P\wedge(R-S)\}^2+\{P\cdot(R+S)\}^2}
.\label{mu}\end{gather} Then, a vector $\mu P$ is homogeneous in
$P,$ depends on one-dimensional ray span by non-zero vector $P,$
denoted by $P$-ray.\end{dassu}

\begin{dthm}[Main: isometry-links]\label{main} Consider the isometry
$L\in O_g$ parameterized by simple bivector \eqref{isom}. Then, each
solution of the isometry-link equation, $LR=S,$ for the given
vectors $R$ and $S,$ is given by a $P$-ray, also referred to as the
privileged (exterior) or preferred $P$-ray, subject to conditions
\eqref{condition}-\eqref{mu}. All isometry-links from $R$ to $S,$
are parameterized by variable rays $\{\mu P\},$ and have the
following explicit form,
\begin{multline}L_{(\mu P)\wedge(R-S)}=\id-\frac{2P\otimes\left\{(R-S)^2
gP-2(P\cdot R)g(R-S)\right\}}{P^2(R-S)^2+4(P\cdot R)(P\cdot S)}\\
-\frac{(R-S)\otimes\left\{2P^2g(R-S)+4(P\cdot S)gP\right\}}
{P^2(R-S)^2+4(P\cdot R)(P\cdot S)},\label{isolink}\end{multline}
\begin{gather}L_{(\mu P)\wedge(R-S)}R=S,\qquad \gamma+1=\mu P\cdot(R+S),
\label{isolink2}\\
L_{(\mu P)\wedge(R-S)}S=2(\mu P\cdot S)S+(1-2\mu P\cdot
S)R-(R-S)^2\mu P.\label{isolink3}\end{gather}\end{dthm}
\begin{proof} Let unknown isometry for unknown simple bivector $P\wedge Q,$
is in the form \eqref{s5}-\eqref{isom},
\begin{gather}L_{P\wedge Q}=\id-P\otimes\alpha-Q\otimes\beta,\notag\\
(\gamma+1)\alpha\equiv Q^2gP-(P\cdot Q+\gamma+1)gQ,\qquad
(\gamma+1)\beta\equiv P^2gQ-(P\cdot
Q-\gamma+1)gP.\label{isom1}\end{gather} The link equation, $LR=S,$
imply the vanishing of the tri-vector
\begin{gather}(P\wedge Q)\wedge(R-S)=0\quad\Longleftrightarrow\quad P\wedge
Q=(\mu P)\wedge(R-S)\neq 0.\end{gather} Inserting an ansatz
$Q=\mu(R-S)$ into link equation $LR=S,$ we get $\alpha R=0$ and
$\mu\beta R=1.$ Therefore \eqref{isom1} imply
\begin{align}\alpha R=0\quad&\Longleftrightarrow\quad\gamma+1=\mu
P(R+S),\\\mu\beta R=1\quad&\Longleftrightarrow\quad
\mu=\;\eqref{mu}.\end{align} In order to finish the proof we must
check that the above solution is compatible with the necessary
condition \eqref{s3}.

Instead of the above deduction of \eqref{isolink}, Reader can check
directly that for this explicit isometry $P$-link \eqref{isolink},
the identities \eqref{isolink2}-\eqref{isolink3}, holds.

The isometry-link \eqref{isolink} is the most general complete
solution subject to Assumption \ref{assum}.\end{proof}

Note that in the above $P$-link, a simple bivector, $(\mu
P)\wedge(R-S),$ is up-to $P$-ray $\mu P,$ in some analogy to the
link-solution for single vector \eqref{refl2}, $\mu(R-S),$ that is
up to arbitrary non-zero scalar $\mu.$

\begin{dnote}[Non-uniqueness of isometry-link]
The link equation is crucial for the interpretation of the isometry
$L=L(R,S)$ in terms of the known initial vector $R$ and known final
vector $S.$ In the case of the Lorentz transformations in the
special relativity, the Lorentz-link problem was considered by
Donald Fahnline [Fahnline 1982, Section III], and by van Wyk in
1986. We will consider the specific pure Lorentz transformations in
details in the next Section.

The non-uniqueness of the isometry-link \eqref{isolink} is important
for understanding the isometry, and for understanding the Einstein
special relativity. We can say that \eqref{isolink}-\eqref{isolink2}
is the link from $R$ to $S$ as \textit{seen} by `a privileged =
preferred' vector $P.$ We abbreviate this isometry-link
\eqref{isolink} briefly by $P$-link.

For \textit{each} vector $P$ subject to conditions
\eqref{condition}, this isometry-$P$-link is a `pure'
transformation, in the meaning used for pure Lorentz transformations
in special relativity, \ie for transformations that do not involve a
spatial rotation. $P$-dependent isometry-link $L_{(\mu
P)\wedge(R-S)},$ \eqref{isolink}, for $R=S$ is an identity,
$L|(R=S)=\id.$ The non-uniqueness of the isometry-link due to $(\mu
P)$-dependence, \eqref{isolink}, has nothing to do with Wigner's
little groups.

This technical non-uniqueness is equivalent to non-uniqueness of the
group embedding $O(3)\;\hookrightarrow\; O(1,3).$

We are going to demonstrate in the next Sections, and in more
details in the separate publication, that exactly the
preferred-vector-dependence of the isometry-link, \ie
non-uniqueness, is the primary source of the non-associativity of
the velocity addition in special relativity [Ungar 1990, 2001], the
primary source of the Thomas precession = rotation = gyration
[Fisher 1972; Ungar 1988, 1989, 1991, 1997, 2001; Urbantke 1990],
the primary source that the relativistic dynamics of many-body
system in special relativity can not be resolved, the primary source
of some problems in electromagnetism [Valent 2002], the primary
source of some paradoxes of the special relativity, e.g. [Sastry
1987], including the Mocanu paradox [1986], and the primary source
that the special relativity is miss-understood, etc.\end{dnote}

The most important conclusion at this point is that the above
bivector-parametrization of isometry,
\eqref{mp}-\eqref{isom}-\eqref{W}-\eqref{isolink}, holds for
dimension no less than two, and for arbitrary signature of the
metric tensor.

\begin{dthm} The link equation $L_{P\wedge(R-S)}R=S$ for
the case when $(R-S)^2=0,$ for unknown vector field $P$ and scalar
field $\gamma$ is reduced to the following set of two scalar
conditions
\begin{gather}(\gamma+1)(P\cdot R-1)=(P\cdot R)(P\cdot R-P\cdot S),\qquad
\gamma^2=1+(P\cdot R-P\cdot S)^2.\end{gather}\end{dthm}

\section{Planar system of three vectors}
\begin{ddef}[Planar system] A system of three vector fields
$\{P,R,S\},$ is said to be planar if vanishes the tri-vector field,
\begin{gather}P\wedge R\wedge S=0.\end{gather}\end{ddef}
The cases of planar and no-planar preferred-vector $P,$
\textit{relative} to the given plane $R\wedge S,$ \ie relative to
the initial $R$ and final $S$ vectors, are essentially different,
and deserve the separate consideration.

From now on, the vectors $R$ and $S$ are such that $R^2=S^2\neq 0.$
Then we have idempotents \eqref{idem},
\begin{gather}r\equiv\frac{R\otimes gR}{R^2}\quad\text{and}\quad s\equiv
\frac{S\otimes gS}{S^2},\quad(\id-r)R=0.\label{idem0}\end{gather}

\begin{dlem}[Planar ternary system] Let $R^2=S^2\neq 0\quad
\text{and}\quad(S+R)^2\neq 0.$ Then\begin{gather}\{(\mu
P)\wedge(R-S)\}|_{P\wedge R\wedge S=0}=\frac{S\wedge
R}{S^2}.\label{Matol0}\end{gather} When in Theorem \ref{main}, a
preferred $P$-ray is coplanar relative to the given plane $R\wedge
S,$ then the bivector generating the pure-isometry-link is given
uniquely in terms of initial and final vectors.\end{dlem}

The planar $g$-isometry-link, $L_{\frac{S\wedge R}{S^2}},$ from $R$
to $S,$ has the following expression
\begin{gather}L_{\frac{S\wedge R}{R^2}}=\id-2\frac{(R+S)\otimes g(R+S)}
{(R+S)^2}+2\frac{S\otimes R}{S^2},\label{Matol}\\
L_{\frac{S\wedge R}{R^2}}R=S\qquad\text{and}\qquad L_{\frac{S\wedge
R}{R^2}}S=(2s-\id)R.\label{Matol2}\end{gather} The planar
$g$-isometry-link \eqref{Matol}-\eqref{Matol2}, generalize for any
dimensions $\geq 2,$ and arbitrary signature, the isometry derived
by Fahnline [Fahnline 1982, Section III, formulas (15)-(16)-(18)],
see \eqref{Fahnline}.

The link-equation $LR=S$ together with the extra condition
\eqref{Matol2}, is equivalent to the Matolcsi definition [Matolsci
1994, \S 1.3.8], and fixes the preferred $P$-ray in
\eqref{isolink}-\eqref{isolink3}, uniquely to be planar,
\begin{gather}\left.\begin{array}{l}LR=S\\LS=(2s-\id)R\end{array}
\right\}\quad\Longleftrightarrow\quad P\wedge R\wedge
S=0.\end{gather}

It is true that the isometry $L_{P\wedge Q}$ is uniquely given by
bivector $P\wedge Q.$ However it is \textit{not} true that the
$g$-isometry-link, $L_{P\wedge Q},$ is uniquely given by initial $R$
and final $S$ vectors, as could be suggested e.g. in [Ungar 2001,
page 348, Theorem 11.16]. The bivector $P\wedge Q,$ generating the
isometry, is not uniquely given by initial $R$ and final $S$
vectors! To get unique isometry-link one need to chose a preferred
$P$-ray \eqref{condition}, or equivalently, put some extra condition
on isometry, like the Fahnline condition \eqref{Matol2} [Fahnline
1982, Section III], assumed explicitly also by Matolcsi [1994, \S
1.3.8], and assumed implicitly in [Ungar 1988, 2001].

Each $P$-link from $R$ to $S,$ is given in terms of the simple
bivector, and therefore each must have the same third order minimal
polynomial \eqref{mp} uniquely determined by the scalar magnitude of
the bivector. Therefore one-way to see for the given pair of
vectors, initial $R$ and final $S,$ that non-planar $P$-link
\eqref{isolink} is essentially different from planar-link
\eqref{Matol}, is to verify the difference of the scalar magnitudes
\begin{gather}P\wedge R\wedge S\neq 0\quad\Longleftrightarrow\quad\{(\mu P)
\wedge(R-S)\}^2\quad\neq\quad\frac{(S\wedge R)^2}{S^4}
\end{gather}

The planar-link \eqref{Matol}, gives the link-identity
\eqref{Matol2}. However we must avoid incorrect impression of the
uniqueness of the Lorentz-link, and not forget about non-planar
solutions of the link-equation \eqref{isolink3}, for dimensions
$\geq 3.$

The planar isometry-link, $R\longmapsto S,$ is generated by bivector
$\frac{S\wedge R}{S^2=R^2},$ however must not be tempted to be
defined as the \textit{only} unique isometry-link. Compare for
example how this subject is presented when specialized for the
four-dimensional Minkowski space-time with the metric tensor of
Lorenztian signature. Among many other we refer to [Fahnline 1982,
Section III, formulas (15)-(16)-(18); Matolcsi 1994, \S 1.3.8; Ungar
2001, Theorem 11.16 on page 348 does not holds for non-planar
preferred observer; Urbantke 2003]. Fahnline [1982 Section III]
gives isometry-link \eqref{Matol}-\eqref{Matol2} for irrelevant
`inertial frame'=basis, and comment that `there are other (pure)
Lorentz transformations'-links from $R$ to $S$ (the other are
non-planar).

The happy conclusion, that the pure isometry-link, $R\rightarrow S,$
is \textit{uniquely} given by initial $R$ and final $S$ vectors,
[Ungar, Theorem 11.16 on page 348], is incorrect conclusion for
dimensions $\geq 3.$ Such conclusion ignore the multitude of
non-planar $g$-isometry-links from initial $R$ to final $S$ given by
\eqref{isolink}. To have just one Lorentz boost from $R$ to $S$ you
need made \textit{a choice} of a preferred $P$-ray
\eqref{condition}, to be planar or non-planar, with many
possibilities for non-planarity relative to the given plane $R\wedge
S.$

The choice of $P$-ray is equivalent to a \textit{choice} of an
embedding $O(3)\;\hookrightarrow\;O(1,3).$

\subsection{Why the uniqueness of the boost is so much desired?} Let me
comment at this moment about Fahnline's expression for the
isometry-link \eqref{Matol}-\eqref{Matol2}, that assume implicitly
the planar preferred-ray. Fahnline arrived to his expression
starting from the pure Lorentz transformation of coordinates,
parameterized by `ordinary relative velocity', following Einstein
[1905] (we consider this in Sections VI-VII). The Fahnline planar
isometry-link is equivalent to Einstein's innocent coordinate
transformation, that hide the physical meaning of the `ordinary
velocity' among reference systems. The pure Lorentz transformation
are parameterized by `ordinary relative velocity' only, having the
most important physical interpretation, and therefore, since 1905,
there is absolute certainty, dogma, about one-to-one correspondence
among relative velocity, that must be \textit{unique} among two
reference systems, and the pure Lorentz transformations-links.
Shortly: the unique velocity of reference system $S$ relative to
reference system $R,$ must imply the unique Fahnline's isometry-link
\eqref{Matol}. Many different non-planar preferred-ray-dependent
pure Lorentz transformations-\textit{links} from $R$ to $S,$ as
given by \eqref{isolink}, would destroy happiness because imply
non-unique relative velocity among reference systems, that was
\textit{not} Einstein's intention in 1905. This is, I believe, the
psychological reasons that during XX century the physics community
accepted the Einstein coordinate transformations, equivalent to
Fahnline's \textit{planar} pure Lorentz transformation-link
\eqref{Matol}-\eqref{Matol2}), as the \textit{unique} boost. This
ignore Theorem \ref{main}, giving the infinite family of non-planar
pure Lorentz transformations-links, with infinite family of
non-planar relative isometric velocities among two bodies.

The isometric velocity, parameterizing the pure Lorentz
transformation, needs, besides these two reference systems
considered by Einstein in 1905, also the obligatory choice of the
planar or non-planar preferred $P$-ray. Such relative velocity we
call, in what follows, the \textit{ternary} velocity,
\begin{gather}\vel=\vel(\text{privileged=preferred}\,P,\;\text{initial}\,R,\;
\text{final}\,S).\label{ternary}\end{gather}

If the unique solution of the link equation would be planar only (as
it is for dim = 2, and the wishful-thinking for dimensions $\geq 3$
[Ungar 2001]), therefore, we would be in heavens. Everything in
special relativity would be crystal clear: unique boost imply unique
(Einstein's reciprocal) relative velocity among reference systems
$R$ and $S,$ and uniqueness of the relative velocity must imply the
associative addition of velocities. Whereas Ungar discovered in 1988
[p. 71] that addition of Einstein `unique velocities' is
\textit{non}-associative.

\begin{dnote} Van Wyk in 1986, used implicitly the bivector-parametrization,
and concludes that even a given \textit{pair} of arbitrary initial
vectors and a given pair of arbitrary final vectors do \textit{not}
provide enough information to specify unique Lorentz-link-boost. In
spite of this, in [Ungar 2001, Theorem 11.16 on page 348], there is
a suggestion (incorrect) that the Lorentz boost-link in four
spacetime dimensions is uniquely given by single initial vector and
single final vector.\end{dnote}

\section{Orthogonal presentation} The $P$-link, from the given
vector $R$ to the given vector $S,$ as given by Theorem \ref{main},
has more transparent interpretation in the orthogonal presentation
of the simple bivector as discussed in Subsection \ref{V-presen}.

The planar-link \eqref{Matol}-\eqref{Matol2}, is given uniquely in
terms of the initial vector $R$ and the final vector $S.$ The
bivector \eqref{Matol0}, generating the planar-isometry-link, has
the following presentations
\begin{gather}\frac{S\wedge R}{S^2}=\{(\id-r)(S-R)\}\wedge\frac{R}{R^2}=
\frac{S}{S^2}\wedge\{(\id-s)(R-S)\}.\end{gather} If additionally
$R\cdot S\neq 0,$ then it is convenient for the future discussion to
introduce the following vector that is non-skew-symmetric linear
span of $R$ and $S,$ \ie it is `non-Galilean subtraction',
\begin{gather}\varpi(R,S)\equiv\frac{R^2}{R\cdot S}(\id-r)(S-R)\neq-\varpi(S,R),
\quad R\cdot\varpi(R,S)=0,\label{binary}\\\{\varpi(R,S)\}^2=-R^2+
\frac{R^6}{(R\cdot S)^2}, \quad(R\cdot
S)^2=\frac{R^6}{R^2+\{\varpi(R,S)\}^2},\label{binary2}\\
\frac{1}{R^2}S\wedge R=\frac{1}{R^4}R\wedge(-R\cdot
S)\varpi(R,S).\end{gather}

Note that the scalar magnitude of the above binary vector
\eqref{binary}-\eqref{binary2}, is symmetric relative to the
exchange $R\leftrightarrow S.$ The vector field (vector)
\eqref{binary}-\eqref{binary2} generalize the concept of the binary
relative velocity, introduced in [\'Swierk 1988; Matolcsi 1994, \S
4.3; Bini 1995; Gottlieb 1996]. The expression \eqref{binary} we
will interpret later on in the four-dimensions that `the velocity
$\varpi(R,S)$ of $S$ relative $R$' is $g$-perpendicular to
`4-vector-velocity $R$'. With this respect it is interesting to
compare with what is emphasized by Cui in interesting paper [Cui
2006], that instead of the relative velocity, the force $\simeq$
acceleration should be $g$-orthogonal to an observer $R,$ as it is
exemplified by the Lorentz force in electromagnetic field.

For arbitrary non-planar $P$-link $R\mapsto S,$
\eqref{isolink}-\eqref{isolink2}-\eqref{isolink3}, the simple
bivector generating isometry-link, is manifestly reciprocal (exactly
skew-symmetric) relative to the exchange $R\leftrightarrow S,$
\begin{multline}(\mu P)\wedge(R-S)=\mu P\wedge\{(\id-p)(R-S)\}\\
=\mu\left\{\left(\id-\frac{(R-S)\otimes g(R-S)}{(R-S)^2}
\right)P\right\}\wedge(R-S),\label{V1}\end{multline}
\begin{gather}\{(\id-p)(R-S)\}^2=\frac{1}{P^2}(P\wedge(R-S))^2.\end{gather}
In terms of the ternary vector $W=W(P,R,S)\equiv\mu(\id-p)(R-S),$
the arbitrary $P$-link $R\mapsto S,$ has the explicit form given by
\eqref{W}, and these expressions \eqref{W}, now must be supplemented
by the following $P$-link identity
\begin{gather}L_{P\wedge W}R=S,\quad W|_{P\wedge R\wedge S=0}=
\frac{4(R+S)^2}{(R+S)^4+4(S\wedge R)^2}\frac{P}{P^2}\cdot(S\wedge
R).\end{gather}

We stress again that all results of entire all previous Sections
holds for dimensions of the vector space (or module) $\geq 2,$ and
for arbitrary signature of the invertible symmetric metric tensor
$g$ (tensor field).

\section{The Einstein reciprocal relative velocity}
In this Section we consider four-dimensional space-time with the
Lorentzian signature of the metric tensor $g,$ $(-+++).$ The main
results of the previous Sections will be specified to this
signature.

What could be a category of the reference systems in the Einstein
and Minkowski relativity? An object of a category of reference
systems can be either time-like monad, or, a tetrad=frame,
$\tau\varepsilon\tau\rho\alpha\delta o\sigma,$ the Lorentz
ortho-normal (1+1+1+1)-split, the Lorentz frame=basis, as it is for
example in [Misner, Thorne, Wheeler 1973], where, following
Einstein, the coordinate system, and the reference system given by
coordinate-free tetrad, is not always distinguished. Every morphism,
and each change of a reference system, is the Lorentz isometric
transformation. The relative velocity among reference systems is
defined in terms of the isometric Lorentz-boost, as a (not unique,
$P$-seen) boost-linking two given time-like `four'-vectors, see
\eqref{isolink}.

\begin{dnote}[Fahnline 1982] Robertson and Noonan [1968, p. 66], and Fahnline
[1982], reexpressed the pure Lorentz coordinate transformation, \ie
the boost parameterized by `velocity of one reference frame relative
to another', as was presented by Einstein in 1905, in terms of the
initial time-like reference vector $R,$ and the final time-like
reference vector $S.$

When the vectors $R$ and $S$ are both time-like, $R^2=S^2=-1,$ and
future-directed, then $R\cdot S\leq -1.$ In this case in formula
\eqref{condi}: $\gamma+1=1-R\cdot S.$ Then, the isometry $L_{R\wedge
S}$ \eqref{isom}-\eqref{Matol} coincide with the Fahnline expression
of pure Lorentz transformation (\ie the boost), [Fahnline 1982;
Matolcsi 1993, 1.3.8; Matolcsi and Goher 2001 formula (27) on p.
91],
\begin{gather}L_{R\wedge S}=\id-2\,S\otimes gR+\frac{(R+S)\otimes g(R+S)}
{1-R\cdot S}.\label{Fahnline}\end{gather} The above expression for
$g$-isometry \eqref{Fahnline}, Fahnline derived from the Lorentz
coordinate transformation.\end{dnote}

The Lorentz boost is known variously but synonymously as: rotation
on imaginary angle [Sommerfeld 1909], non-rotational Lorentz
transformation [Fisher 1972], pure Lorentz transformation [Fahnline
1982]. In what follows we will restrict the name Lorentz boost for
isometry-link parameterized explicitly in terms of the
\textit{relative velocity.} For this we need to define what means a
`relative velocity' in connection with basis-free isometry-link?

\begin{ddef}[Isometric velocity] The relative velocity parameterizing
the isometry, $L(\vel)\in O_g,$ $L(\vel=\zel)\equiv\id,$ is said to
be \textit{isometric}-velocity, or, the Einstein velocity. The
isometric-velocity must be reciprocal
$\{L(\vel)\}^{-1}=L(-\vel),$\begin{gather}L(\vel)R=S\qquad\Longleftrightarrow
\qquad L(-\vel)S=R.\end{gather}\end{ddef}

Reader interested in special relativity of Albert Einstein [Einstein
1905], have the right to ask, why bivector parametrization,
\eqref{mp}-\eqref{isom}-\eqref{W}-\eqref{isolink}, is important?
Following Lorentz [1904], Poincar\'e [1904], and Einstein [1905],
the textbooks parameterizes Lorentz boost in terms of a single
vector `of relative velocity' that it is suggested to be outside of
the Minkowski spacetime, in some idealized absolute Euclidean space.
This assume that exists 'standard' embedding, given by heavens,
$O(3)\quad\hookrightarrow\quad O(1,3).$

However, all observed relative velocities are tangent to
simultaneity of observer, assuming that an observer can not see
moved bodies in his own past or in his own future. Therefore
relative velocities \textit{are} space-like vectors \textit{in}
Minkowski space-time. A single vector can parameterize an isometry
if an isometry $L$ is a reflection = unipotent
\eqref{refl1}-\eqref{refl2}, or, if we chose the preferred embedding
$O(3)\;\hookrightarrow\;O(1,3).$ Otherwise, we need at least a
simple bivector, that is relevant for parametrization of
non-unipotent isometry. Many Authors consider the matrix
representation of the Lorentz boost as `the best definition', for
example in [Ungar 1988 page 62; 2001, page 254]. In fact it is just
contrary, the matrix representation, irrelevant-basis-dependent,
introduce irrelevant information, whereas the relevant bivector is
lost or it is hidden, and therefore matrix representation does not
elucidate the correct meaning of the Einstein `relative velocity'
parameterizing Lorentz isometry. Our aim is to show the place of
this simple bivector in Lorentz-boost transformations.

Bivectors are relevant for \textit{every} isometry. For the
\textit{Lorentz} isometry group, the relevance of the bivectors  was
emphasized from another perspective by Ehlers, Rindler and Robinson
[1966], [Rindler and Robinson 1999], and by Baylis and Sobczyk
[2004].

A massive observer, a reference system, we identify with time-like
future-directed and normalized vector fields, $P^2=-1.$ The
simultaneity of an observer $P$ is given by $g$-dependent
differential Pfaffian one-form $-gP.$ All observers by definition
are $g$-orthogonal in this meaning, all observers are orthogonal
projectors, \ie are orthogonal idempotents.

For Lorentzian signature, if a vector $P$ is time-like, and $P\cdot
V=0,$ then vector $V$ is space-like.

\begin{ddef}[Observing velocity] For a time like vector $P$ (representing
massive reference system), and a space-like vector $\vel$
(representing the possible relative velocity), the condition
$P\cdot\vel=0,$ is interpreted as the necessary and sufficient for
observing $\vel$ by $P.$\end{ddef}

The orthogonality condition $P\cdot\vel=0,$ means that the vector
$\vel$ is tangent to simultaneity of time-like observer, $P^2=-1.$
We assume that every observer see the relative velocities among
bodies, as tangent to his simultaneity, and not in his past or in
his future.

The Heaviside-FitzGerald-Lorentz scalar factor is denoted by
\begin{gather}\gamma_\vel\equiv\left(1-\frac{\vel^2}{c^2}\right)^{-\half},
\qquad\frac{\vel^2}{c^2}=\frac{\gamma_\vel^2-1}{\gamma_\vel^2},\qquad
\overline{\vel}\equiv\gamma_\vel\frac{\vel}{c},\qquad\overline{\vel}^2=
\gamma_\vel^2-1.\end{gather}

Let's come back to bivector \eqref{V1}, solving the isometry-link
subject to Condition \ref{assum},
\begin{gather}L_{(\mu P)\wedge(R-S)}R=S,\qquad
(\mu P)\wedge(R-S)=P\wedge\mu(\id-p)(R-S).\end{gather}

\begin{ddef}[Ternary isometric relative velocity] Let the system of three
time-like vectors (vector fields) $\{P,R,S\}$ represents the
three-body massive system, \ie $$P^2=R^2=S^2=-1.$$ The ternary
isometric velocity $\vel=\vel(P,R,S)$ of a massive body $S$ relative
to massive body $R,$ as observed by a massive reference system $P,$
is defined as follows
\begin{gather}\overline{\vel}\equiv\gamma_\vel\frac{\vel}{c}\equiv\mu(\id-p)
(R-S)|_{P^2=R^2=S^2=-1},\label{ternary1}\\
L_{P\wedge\overline{\vel}}R=S\quad\Longrightarrow\quad\text{unique}
\quad\overline{\vel}=\mu(\id-p)(R-S).\label{ternary0}\end{gather}\end{ddef}

\begin{olem} The following formulas holds
\begin{gather}\gamma_{P\wedge\overline{\vel}}=\gamma_\vel,\quad
\gamma_{(\mu P)\wedge(R-S)}=\left|\frac{\{P\cdot(R+S)\}^2-\{P\wedge(R-S)\}^2}
{\{P\cdot(R+S)\}^2+\{P\wedge(R-S)\}^2}\right|,\\
\gamma_{(\mu P)\wedge(R-S)}|_{P\wedge R\wedge S=0}=|R\cdot S|,\\
\overline{\vel}|_{P\wedge R\wedge S=0}=
\frac{4(R+S)^2}{(R+S)^4+4(S\wedge R)^2}\frac{P}{P^2}\cdot(S\wedge
R)=P\cdot(S\wedge R).\end{gather}\end{olem}

The isometric velocity $\vel=\vel(P,R,S)$ \eqref{ternary0} is
$P$-dependent velocity of $S$ relative to the reference system $R.$

All vector fields (vectors) of the system $\{P,R,S\}$ must be
time-like normalized in order that the bivector, $(\mu
P)\wedge(R-S),$ generating an isometry $R\mapsto S,$ can be
interpreted in terms of the physical relative velocity
$\vel=\vel(P,R,S)$ among massive bodies. Note that no one time-like
vector field in the ternary system $\{P,R,S\}$ needs to be inertial.

Is evident that the Einstein relative isometric planar or no-planar
relative ternary velocity, $\vel(P,R,S),$ is not given uniquely by
the reference systems $R$ and $S,$ because needs the choice of the
preferred observer $P$ in Definition \eqref{ternary0} and in Theorem
\eqref{Lorentz0}. The Einstein velocity depends on the choice of the
\textit{preferred} massive observed $P.$ One can say that the
isometric and reciprocal velocity of $S$ relative to $R$ is
\textit{seen} by preferred observer $P.$ This is why the relative
isometric-velocity \eqref{ternary1} is said to be the
\textit{ternary velocity}. This is a velocity of a reference system
$S$ relative to $R$ as \textit{seen} by $P.$ For the fixed reference
systems, $R$ and $S,$ and the variable choice of the preferred
observer $P,$ there could be infinite many Einstein's velocities of
$S$ relative to $R,$ as \textit{seen} by many different privileged
variable observers $P.$

\begin{ddef}[Lorentz boost] Let $P^2=-1$ and $P\cdot\vel=0.$ Let
introduce the following (differential) forms,
\begin{gather}\nu\equiv(\gamma_\vel-1)gP-g\overline{\vel},\qquad
\xi\equiv gP-\frac{g\overline{\vel}}{\gamma_\vel+1}.\end{gather}

A basis-free boost is a pure Lorentz isometry transformation
parameterized in terms of the bounded space-like velocity $\vel$ in
terms of the bivector $P\wedge\gamma_\vel\vel/c,$ as follows
\begin{multline}L_{P\wedge\overline{\vel}}=\id-P\otimes\nu-\overline{\vel}
\otimes\xi\\=\id-(\gamma_\vel-1)P\otimes
gP+\gamma_\vel\left(P\otimes g\frac{\vel}{c}-\frac{\vel}{c}\otimes
gP\right)+\frac{\gamma_\vel^2}{\gamma_\vel+1}\,\frac{\vel\otimes
g\vel}{c^2},\label{W2}\end{multline}\end{ddef}

For the given space-like bounded velocity $\vel,$ there is a
two-dimensional manifold of allowed time-like massive
$\vel$-observers \{$P$\} in \eqref{W2} (observer of $\vel$ is
\textit{not} unique), because in dimension four the set of two
conditions, $P\cdot\vel=0$ and $P^2=-1,$ has the two-parameter
family of solutions. The observer-\textit{dependent} Lorentz boost
\eqref{W2} must be compared with the frequent definition of the
Lorentz boost as the matrix parameterized 'uniquely' in terms of the
velocity $\vel$ only, see for example in [M{\o}ller 1952, 1972;
Ungar 1988, page 62, formulas (6)-(8); Ungar 2001, page 254,
formulas (8.2)-(10.29)].

From the boost definition \eqref{W2}, the following expressions
follows,
\begin{multline}L_{P\wedge\overline{\vel}}R=R-\left\{(\gamma_\vel-1)
(P\cdot R)-\gamma_\vel\left(\frac{\vel}{c}\cdot R\right)\right\}P-\\
\gamma_\vel\left\{(P\cdot R)
-\frac{\gamma_\vel}{\gamma_\vel+1}\left(\frac{\vel}{c}\cdot
R\right)\right\}\frac{\vel}{c},\end{multline}
\begin{gather}\left\{L_{P\wedge\overline{\vel}}\right\}^{-1}=
L_{P\wedge(-\overline{\vel})},\quad P\wedge R\wedge
\left(L_{P\wedge\overline{\vel}}R\right)=-(\xi R)P\wedge
R\wedge\overline{\vel},\label{noplanar}\\
R\cdot\left(L_{P\wedge\overline{\vel}}R\right)=R^2-
(\gamma_\vel-1)(P\cdot R)^2+\frac{\gamma_\vel}{\gamma_\vel+1}
\left(\frac{\vel}{c}\cdot R\right)^2,\\L_{P\wedge\overline{\vel}}P=
\gamma_\vel\left(P+\frac{\vel}{c}\right).\end{gather}

\section{Lorentz coordinate transformations}
In what follows we assume that $R$ is time-like vector field such
that $R^2=-1$ (a massive observer), with an associated idempotent
operator, $r\equiv R\otimes(-gR).$ Then
$L_{P\wedge\overline{\vel}}R$ is a massive body that moves with the
velocity $\vel$ relative to $R,$ as measured by reference system
$P.$ Let these two observers, $R$ and $LR,$ are observing the
particle=event $e.$ Then with respect to the reference system $r,$
particle $e$ is
\begin{gather}e=re+(\id-r)e=ctR+\xel=ct'LR+\xel',\\
ct\equiv-R\cdot e\quad\text{and}\quad ct'\equiv-(LR)\cdot
e=-R\cdot(L^{-1}e).\end{gather} Here, $R\mapsto LR,$ is isometry
action that change the reference system, and, $e\mapsto L^{-1}e,$ is
the change of the observed coordinated particle.

The massive bodies $R$ and $S\equiv LR,$ understood as the
normalized time-like vector fields, can be identified with the pair
of the reference systems, in mutual motion, considered by Albert
Einstein in 1905.

Then $\xel=\xel(r,e)\equiv(\id-r)e$ is directly observable
coordinate=location of a particle=event $e$ relative to the
reference system $R.$ Correspondingly, $\xel'=\xel(S=LR,e)$ be a
coordinate of the same particle=event $e$ as seen by the reference
system $LR=S.$

We wish to apply the basis-free boost \eqref{W2}, for the
textbooks's case
\begin{gather}L_{P\wedge(-\overline{\vel})}\begin{pmatrix}ct\\\xel\end{pmatrix}=
L_{P\wedge(-\overline{\vel})}(ctR+\xel)
=\begin{pmatrix}ct'\\\xel'\end{pmatrix}=ct'R+\xel'.\end{gather}

How the choice of the preferred time-like observer $P,$ observing
the velocity $\vel,$ $P\cdot \vel=0,$ would appears in Lorentz
transformation of coordinates of particle $e$ observed by $R$ and
$LR$ ?

\begin{dthm}[Lorentz coordinate transformation]\label{Lorentz0} The general
isometric pure Lorentz transformation of coordinates of an event $e$
as seen by the reference systems $R$ and $LR$ respectively, is
parameterized by bivector $P\wedge\overline{\vel},$ [and not by 'a
velocity parameter $\vel$' alone], and depends on preferred observer
$P,$
\begin{gather}\nu e=ct\{(\gamma_\vel-1)P\cdot R+\overline{\vel}\cdot
R\}+\overline{\vel}\cdot\xel+(\gamma_\vel-1)P\cdot\xel,\\
\xi e=ct\left(P\cdot R+\frac{\overline{\vel}\cdot
R}{\gamma_\vel+1}\right)+\frac{\overline{\vel}\cdot
\xel}{\gamma_\vel+1}+P\cdot\xel,\\
ct'=ct+(gR)\{(\nu e)P-(\xi e)\overline{\vel}\},\\
\xel'=\xel-(\id-r)\{(\nu e)P-(\xi e)\overline{\vel}\},\\
(\id-L^{-1})e=(\nu e)P-(\xi e)\overline{\vel}.\end{gather}\end{dthm}

Thus the above Lorentz transformation, when compared with the
traditional presentations [Einstein 1905; Fock 1955, 1959, 1961,
1964 \S 10; Jackson 1962, 1975 \S 11.3, Zachary and Gill, the
present volume], depends on three new scalars:
\begin{itemize}
\item $R\cdot P$ - describe the mutual motion among reference system
$R$ and preferred observer $P.$
\item $R\cdot\vel$ - describe how much the relative velocity
observed by $P$ is far away from simultaneity of the reference
system given by time-like normalized vector field $R.$
\item $P\cdot\xel$ - describe how much the position of an event $e$
relative to the reference system $R,$ is far away from simultaneity
of the preferred observer $P.$
\end{itemize}

In the Gallilean limit of the absolute simultaneity,
$c^2\rightarrow\infty,$ $R\cdot\vel=0,$ Theorem \eqref{Lorentz0}
gives $t'=t,$ and $\xel'=\xel-t\vel.$

The 'singular' case of the coplanar preferred observer, and the
general case of no-planar preferred observer, deserve to be consider
separately, \eqref{noplanar},
\begin{multline}P\wedge R\wedge
\left(L_{P\wedge\overline{\vel}}R\right)=-(\xi R)P\wedge
R\wedge\overline{\vel}\\=\begin{cases}\;0&\text{for coplanar
preferred observer $P$}\\\neq 0&\text{for non-planar preferred
observer $P.$}\end{cases}\end{multline}

Moreover, for non-planar preferred $P,$ still the particular case
when $P$ and $R$ are in mutual motion $P\cdot R\neq-1,$ however both
can observe the \textit{same} velocity $R\cdot\vel=0=P\cdot\vel,$
also deserve the separate consideration. We will consider the entire
analysis of general non-planar Lorentz transformation in a separate
publication.

Let a preferred observer $P$ be planar relative to the plane
$R\wedge LR,$
\begin{gather}P\wedge R\wedge
\left(L_{P\wedge\overline{\vel}}R\right)=-(\xi R)P\wedge
R\wedge\overline{\vel}=0.\end{gather} If $\xi R\neq 0$ and
$(R\cdot\overline{\vel})^2+\gamma_\vel-1\neq 0,$ then the Lorentz
transformation is parameterized by two additional scalar parameters,
$R\cdot P$ and $R\cdot\vel,$ only, because there is the following
identity,
\begin{gather}\{(R\cdot\overline{\vel})^2+\gamma_\vel^2-1\}P\cdot\xel=
(P\cdot R)(R\cdot\overline{\vel})(\xel\cdot\overline{\vel}).
\end{gather}

\begin{dex}[Einstein 1905; Fock 1955, 1959, 1961, 1964 \S 10; Jackson 1962,
1975 \S 11.3] Let $P=R.$ Then all three additional scalars are
fixed: $(\id-r)P=0,$ $P\cdot R=-1,$ $R\cdot\vel=0,$ and
$P\cdot\xel=0.$ Therefore $\vel$ is a velocity of a reference system
$S=LR$ relative to reference system $R$ as \textit{seen} by $P=R.$
The Lorentz-boost isometry-transformation of coordinates of an event
$e,$ is as follows
\begin{gather}\xel'=\xel+\frac{\gamma_\vel^2}{\gamma_\vel+1}\frac{(\vel\cdot\xel)}
{c^2}\vel-\gamma_\vel\vel t,\quad
t'=\gamma_\vel\left(t-\frac{\vel\cdot\xel}{c^2}\right),\label{lt}\\
\frac{\vel\cdot\xel}{c^2}=t-\frac{t'}{\gamma_\vel},\qquad
\vel=\left(\frac{\gamma_\vel+1}{\gamma_\vel}\right)
\left(\frac{\xel-\xel'}{t+t'}\right),\qquad
\gamma_\vel=\frac{1+\frac{1}{c^2}\left(\frac{\xel-\xel'}{t+t'}\right)^2}
{1-\frac{1}{c^2}\left(\frac{\xel-\xel'}{t+t'}\right)^2}\,,\label{Urb}\\
\vel=\vel(R,R,LR)=2\left\{1+\frac{1}{c^2}
\left(\frac{\xel-\xel'}{t+t'}\right)^2\right\}^{-1}
\left(\frac{\xel-\xel'}{t+t'}\right),\label{Urbantke}\\
\vel=\vel(R,R,LR)=-\vel(R,LR,R).\end{gather}\end{dex}

In the Gallilean limit of the absolute simultaneity,
$c^2\rightarrow\infty,$ $t'=t$ and the Urbantke formula
\eqref{Urbantke} gives $\vel=\frac{\xel-\xel'}{t}.$

\section{Non-associative addition of isometric relative velocities}
The explicit coordinate expression \eqref{Urbantke} for the relative
velocity parameterizing the Lorentz coordinate transformation, was
derived, in another way, by Urbantke [2003, p. 115, formula (7)],
and in another form using gyration by Ungar [2001, p. 348, Theorem
11.16]. The Urbantke expression can be interpreted as the
operational \textit{definition} of the Einstein relative velocity in
terms of the directly measurable space-distances and time-intervals
for the case when the preferred observer (Earth) is chosen to be
$S=R,$ compare with the general definition of the ternary relative
velocity \eqref{ternary1}. The Urbantke expression \eqref{Urbantke}
shows that the relative velocity, parameterizing the planar isometry
transformation, must be reciprocal, $\vel^{-1}=-\vel.$ This relative
isometric-velocity is \textit{ternary} velocity because
\textit{assume} that preferred observer is $P=R.$ For the Lorentz
group property, $(L(\vel))^{-1}=L(-\vel),$ the reciprocity of the
velocity is unavoidable, and in fact imply that the
link-transformation $R\mapsto S$ must be an isometry. In particular,
one can choose in Definition \eqref{ternary1} for example $P=S=LR,$
\ie consider the reference system $S$ to be preferred, however the
reciprocity can not be lost,
\begin{gather}\vel(S,R,S)=-\vel(S,S,R)\neq\-\vel(R,S,R).
\end{gather}

How to interpret the Urbantke formula \eqref{Urb}-\eqref{Urbantke}?

The relative velocity $\vel$ among reference systems for the choice
of the preferred \textit{planar} observer $P+R,$
\eqref{lt}-\eqref{Urbantke}, does not involve the distance among $R$
and system $S.$ Instead, this velocity of $S$ relative to $R,$ needs
the two auxiliary distances, $\xel(R,e)$ and $\xel(S,e),$ among
event $e$ and the both reference systems in question. Therefore one
can expect to interpret this reciprocal velocity $\vel$ of $S$
relative to $R$ as seen by $R,$ as the kind of the `subtraction'
from the actually measured pair of velocities, $\uel\equiv\uel(R,e)$
and $\wel\equiv\uel(S,e),$ of the event $e$ relative to the
reference systems $R$ and $S,$ correspondingly.

The extra dependence of the Einstein relative velocities on
preferred observer, \eqref{ternary0}, implies that the addition of
the relative isometric-velocities (Lorentz links among monads or
links among Lorentz frames = bases), is non-associative. The
non-associative addition of Einstein's isometric velocities
[Einstein 1905, Fock 1955], following Vargas [1984, p. 647] and
Mocanu [1986, 1992], is denoted by a symbol $\oplus.$

The appropriate subtraction of relative velocities, giving $\vel$ in
terms of $\uel$ and $\wel,$ is equivalent, implicitly, to the
addition of velocities, expressing for example
$\wel=\uel\oplus(-\vel),$ as the addition of $-\vel$ with $\uel.$ We
will use the following identity
\begin{gather}(\vel\cdot\uel)\vel=\vel\cdot(\uel\wedge\vel)+c^2
\left(1-\frac{1}{\gamma_\vel^2}\right)\uel.\end{gather}

The addition of relative reciprocal isometric-velocities was derived
by Einstein in 1905 using composition of the Lorentz coordinate
transformations (assuming that composed velocities are constant).
The same addition formula was derived by Vladimir Fock using the
Lorentz transformation of the differentials of coordinates
\eqref{lt}, the first differential prolongation of \eqref{lt}, and
assuming the constant isometric relative velocity,
$\vel=\text{const,}$ $d\vel=0,$ only. The equivalent way to derive
the addition $\oplus$ of relative velocities is by means of the
differential prolongation of the formula \eqref{Urb} (instead of
\eqref{lt}),
\begin{gather}d\vel=0\quad\Longleftrightarrow\quad d\left(\frac{\xel-\xel'}
{t+t'}\right)=0,\qquad d\xel=\uel\,dt\qquad\text{and}\qquad
d\xel'=\wel\,dt',\end{gather}
\begin{align}\wel=\uel\oplus(-\vel)&=\frac{\uel-\gamma_\vel\vel}
{\gamma_\vel\left(1-\frac{\vel\cdot\uel}{c^2}\right)}
+\frac{\gamma_\vel}{\gamma_\vel+1}\,\frac{(\vel\cdot\uel)\vel}
{(c^2-\vel\cdot\uel)},\label{Fock3}\\
&=\frac{\uel-\vel}{\left(1-\frac{\vel\cdot\uel}{c^2}\right)}+
\frac{\gamma_\vel}{\gamma_\vel+1}\,\frac{\vel\cdot(\uel\wedge\vel)}
{(c^2-\vel\cdot\uel)},\qquad
\wel\wedge\uel\wedge\vel=0.\label{Fock4}\end{align}

The Einstein-Fock addition of velocities \eqref{Fock3} is an
implicit expression of the Lorentz parametria $\vel$ in terms of
relative velocities $\uel$ and $\wel.$ The explicit subtraction one
can derive using the following naive algebraic definition of
relative velocities $\uel$ and $\wel,$ without differentials, like
e.g. in [Ungar 2001, Theorem 11.16 on page 348],
\begin{gather}\xel=\uel\,t\qquad\text{and}\qquad\xel'=\wel\,t'.\end{gather}
Inserting these definitions into interval, $e^2=(Le)^2,$
$-(ct')^2+{\xel'}^2=-(ct)^2+\xel^2,$ we get the reciprocal velocity
$\vel$ as the subtraction of $\wel$ from $\uel,$
\begin{gather}(t'\,\gamma_\uel)^2=(t\,\gamma_\wel)^2,\qquad
\gamma_\vel\equiv\frac{(\gamma_\uel+\gamma_\wel)^2+\frac{1}{c^2}
(\gamma_\uel\uel-\gamma_\wel\wel)^2}
{(\gamma_\uel+\gamma_\wel)^2-\frac{1}{c^2}(\gamma_\uel\uel-\gamma_\wel\wel)^2},\\
\frac{\gamma_\vel\vel}{\gamma_\vel+1}\equiv\frac{\gamma_\uel\uel-\gamma_\wel\wel}
{\gamma_\uel+\gamma_\wel}.\end{gather} The above expression one can
compare with the Ungar Theorem 11.16, where it is assumed implicitly
that preferred observer is coplanar $P=R.$

We need to stress that all three relative velocities, $\vel,\uel,$
and $\wel,$ entering the addition law \eqref{Fock3}-\eqref{Fock4},
and entering the subtraction operation, are reciprocal, and are
ternary velocities as \textit{seen} by $P,$
$P\cdot\vel=P\cdot\uel=P\cdot\wel=0,$ all these velocities are
tangent to simultaneity of $P.$ Elsewhere we derive the same
addition law of the reciprocal velocities
\eqref{Fock3}-\eqref{Fock4}, without assuming that the relative
velocity $\vel$ is constant.

\subsection{Second differential prolongation} The second differential
prolongation of the addition of velocities \eqref{Fock3} gives
\begin{gather}d\uel=\ael\,dt\quad\text{and}\quad d\wel=\ael'\,dt',\\
\gamma_\vel^2\left(1-\frac{\vel\cdot\uel}{c^2}\right)^2\,\ael'=\ael+
\frac{\vel\cdot\ael}{c^2-\vel\cdot\uel}\left(\uel-\frac{\gamma_\vel\vel}
{\gamma_\vel+1}\right).\label{torr0}\end{gather} There are the
following two identities
\begin{gather}\frac{\gamma_\vel}{\gamma_\vel+1}\{(\vel\cdot\ael)\vel-
\vel\cdot(\ael\wedge\vel)\}
=\left(1-\frac{1}{\gamma_\vel}\right)\ael,\\
\frac{\gamma_\vel^2-1}{\gamma_\vel^2}\{(\vel\cdot\ael)\uel
-(\vel\cdot\uel)\ael\}=(\vel\cdot\uel)\{\vel\cdot(\ael\wedge\vel)\}-
(\vel\cdot\ael)\{\vel\cdot(\uel\wedge\vel)\}.\end{gather}

Torres del Castillo and P\'erez S\'anchez [2006] consider a particle
$e$ accelerated relative to reference systems $R$ and $S.$ Authors
consider the collinear motion only, $\vel\wedge\uel=0,$ and then
\eqref{torr0} gives,
\begin{gather}\gamma_\vel^3\left(1-\frac{\vel\cdot\uel}{c^2}\right)^3\,\ael'
=\ael.\end{gather}

\section{The Einstein principle of reciprocity of relative
velocity} Throughout this paper, the Galilean binary addition of
velocities is denoted by a binary operation symbol `$+$'.

The Einstein formulation of special relativity needs `the Einstein
principle of reciprocity of velocity'. This principle is related to
the identification of the Lorentz-group \textit{parametrization} of
the frames/Koordinatensystem's with the physical inertial observers.
This identification motivates Einstein's reciprocity-axiom [1905, \S
3]:
\begin{gather}\vel\oplus\uel=\zel\quad\Longleftrightarrow\quad\vel+\uel
=\zel\quad\Longrightarrow\quad\vel\wedge\uel=0.\label{E1}\end{gather}
In words: `if the velocity of a frame $S$ relative to $R$ is $\uel,$
then the velocity of $R$ relative to $S$ is $-\uel.$'

In the Einstein reciprocity axiom \eqref{E1} the zero velocity is
the velocity of $R$ with respect to $R,$ $\zel\equiv\zel_R.$
Terletskii [1966, 1968] and Mocanu [1986, 1992] refer to the
Einstein axiom \eqref{E1} as `the Einstein principle of reciprocity
of velocity'.

The Einstein principle of reciprocity is discussed by Cattaneo
[1958], Berzi \& Gorini [1968] and by Newburgh [1972].

The proper-time an exact differential form of an observer is given
by an equivalence relation on events, two events are in the same
equivalence class if they are simultaneous. The proper-time is
relative, is observer-dependent. The Einstein principle \eqref{E1},
that the inverse with respect to the addition of velocities $\oplus$
must be the same as for the Galilean absolute time addition `+',
apparently contradicts the physics of the relativity of
simultaneity. Mutually moved massive bodies/observers/systems must
possess different simultaneous relations. Therefore the relative
velocities measured (in the time-space) by mutually moved observers
must be tangent to the different not co-planar simultaneity
foliations in a time-space.

The relative velocity that obey the Einstein reciprocity must be
preferred observer-dependent, because such relative velocity must be
tangent to simultaneity of the preferred observer, and not
necessarily tangent to simultaneity of observed bodies in mutual
motion.

\begin{dnote}[Speed of the light] The limited velocity of the speed of
light is believed to be the necessary axiom for the special
relativity theory, cf. explicit second axiom in Prinzip der
Relativit\"at, stated by Albert Einstein on the first page of his
fundamental paper [Einstein 1905]: speed of the light, $\cel,$ is
independent on a speed $\vel$ of the radiating massive source,
\begin{gather}\forall\,\vel,\quad(\cel\oplus\vel)^2=\cel^2=
(\vel\,\oplus\,\cel)^2.\label{E0}\end{gather} In fact, \eqref{E0} is
the \textit{consequence} of the law of addition, and the limiting
velocity one can deduce as the theorem from the Einstein-Fock
relativistic law of the addition of relative isometric velocities
\eqref{Fock3}.

We do not agree with the following Adolf Gr\"unbaum opinion,

\begin{quotation} This claim concerning the limiting character of the
speed of light is a fundamental axiom of the theory and is
\textit{not} a theorem depending for its deduction on the
relativistic law for the addition of velocities, as we are sometimes
told. For (i) this addition law itself depends for its own deduction
on the Lorentz transformations \ldots \bf Adolf Gr\"unbaum in
\textit{Logical and Philosophical Foundations of the Special Theory
of Relativity} [1955, 1960 p. 405].\ef\end{quotation}

To get the addition $\oplus$ of relative velocities the Einstein two
explicit axioms (from the first page of the Einstein fundamental
paper) are either not sufficient, nor they are needed at all. We
claim that the addition $\oplus$ of Einstein's isometric velocities
can not be deduced without the reciprocity axiom stated by Einstein
almost at the end of \S 3 of his fundamental paper [Einstein 1905,
\S 3].\end{dnote}

In the relativity theory as the groupoid category, in the
Lorentz-group-free relativity [Oziewicz 2006, Page 2006], there is a
Theorem that the speed of the light is emitter-free for arbitrary
\textit{non}-inertial emitter.

\section{Lorentz-group-free relativity}
In this essay, relativity means relativity of time and relativity of
space, and not yet the gravity theory. In most textbooks the special
relativity means inertial observers and constant relative
velocities, and assume matter-free flat Lorentzian metric tensor
field with no curvature and no torsion, and with Poincar\'e symmetry
group,
\begin{gather}\text{No matter}\quad\Longleftrightarrow\quad\text{no
curvature}\quad\Longleftrightarrow\quad\left\{\text{\bt Empty
spacetime\\with\\Poincar\'e symmetry\et}\right\}\end{gather}

In many respects our presentation follows the ideas of the monograph
by Tom\'as Matolcsi [1994], where coordinate-free (= without
reference frames) relativity theory of spacetime is formulated in
the attractive way that does not need the artificial separation into
special and general relativities [Matolsci 1994, \S 6]. Similarly to
[Matolsci 1994] we consider most general non-inertial observers.
Matolsci define an observer as the normalized time-like vector
field, monad, whereas we identify an observer with trace-class
idempotent, like \eqref{idem0}.

\subsection{Relative velocity as morphism in a connected groupoid
category of null objects} A groupoid is a small category in which
all morphisms are isomorphisms.

The Lorentz-group-free relativity theory holds for arbitrary metric
tensor field and for the most general field of observers, not
inertial, rotated, accelerated, deformed, etc. The
Lorentz-group-free relativity theory, does not needs the concept of
the Lorentz isometric transformations as the relativity
transformations, however does needs some very elementary concepts of
the category theory. The Lorentz-group-free relativity is a
connected groupoid category of massive observers (a connected
groupoid category of massive reference systems), and is denoted by
$\varpi.$

An object of category $\varpi,$ an observer, is
(1+3)-split-idempotent like \eqref{idem0} (in particular, a monad =
$\chi\rho o \nu o\zeta$). A family of objects=observers generate an
associative operator algebra $\Obs(\varpi),$ that most likely is a
Frobenius algebra. Each morphism in this category $\varpi,$ a change
of observer, is given by relative velocity among reference systems.
This is non isometry transformation, so it is not Lorentz
transformation, because the domain is restricted to the sub-manifold
of time-like normalized vector-fields only. The addition of
velocities-morphisms is associative [Oziewicz 2005, 2006, Page
2006].

\begin{table}[h]\caption{The different categories of reference systems.}
\begin{center}\begin{tabular}{c||c|c}\hline&Special relativity&\bt
Lorentz-group-free\\relativity:\\groupoid category\et\\\hline\hline
object&\bt monad or\\Lorentz basis\\$\tau\varepsilon\tau\rho
\alpha\sigma$\et&\bt(1+3)-split\\split-idempotent\\$\chi\rho o \nu
o\zeta$\et\\\hline\bt morphism:\\change\\of observer\et
&\bt isometry:\\Lorentz coordinate\\transformation\et&\bt relative velocity,\\
\textbf{not} isometry\et\\\hline \bt addition of\\relative
velocities\et&\textbf{non}-associative
&associative\\\hline\hline\end{tabular}\end{center}\end{table}

Lorentz-group-free relativity of space and of proper-time does not
need the concepts of isometry and the Lorentz group, nor the
hyperbolic geometry of Lobachevsky [1829] and J\'anos Bolyai [1832],
whereas needs some basic concepts of the category theory [Mac Lane
1998], and need the category of observers, as in [Vladimirov 1982].

The fundamental, primary concept, is the relative non-reciprocal
velocity-morphism between massive split-idempotents.
\begin{gather}\text{No matter}\quad\Longleftrightarrow\quad\text{no
relative velocity}\quad\Longleftrightarrow\quad\text{no
relativity}\end{gather}

The Lorentz group, the symmetry group of the metric tensor for empty
space-time, does not need to be the foundation of the conceptual
relativity theory. Instead, the concept of non-inertial observer
(reference system) seems to be more relevant.

A starting concept is an enriched connected groupoid category of
$(1+3)$-observers/bodies, denoted by $\varpi,$ with the categorical
morphism inter\-pre\-ted as an actual relative velocity, this is the
main axiom. The set of all arrows of a category $\varpi$ with a
given source $p\in\obj\varpi,$ and a given target, $q\in\obj\varpi,$
is abbreviated in the following different ways:
\begin{gather}\varpi(p,q)=\hom_\varpi(p,q)=\hom(p,q)=
\varpi(\text{observer},\text{observed}).\label{hom}
\end{gather} We are using the first notation. The
composition of morphisms reads from right to left,
\begin{gather}\begin{CD}\varpi(q,r)\times\varpi(p,q)\quad @>{\circ}>>\quad
\varpi(p,r),\end{CD}\\
\begin{CD}p\quad @>{\varpi(p,q)}>>\quad q\quad @>{\varpi(q,r)}>>\quad
r.\end{CD}\end{gather} The addition of the velocities-morphisms (not
parameters in the Lorentz boosts) is the associative operation and
is denoted by $\circ$.

If a velocity $\uel$ is a velocity of a body/object $q\in\obj\varpi$
with respect to a laboratory/observer $p\in\obj\varpi,$ we display
(picturing) this velocity $\uel$ as an actual categorical
arrow/morphism/directed-path starting at laboratory/observer/source
$p$ (as a node of the directed graph), and ending at a body/target
$q,$
\begin{gather}\begin{CD}p@>{\uel}>>q\end{CD}\qquad\text{or}\qquad\uel=
\varpi(p,q),\label{display}\\
\text{observer/source/tail/domain of}\;\uel\;=p,\\
\text{observed/target/head/codomain}\;\uel\;=q,\\
\uel=\zel\quad\Longleftrightarrow\quad p\equiv
q,\qquad\zel_p\equiv\varpi(p,p).\end{gather} This reads: the
velocity $\uel$ is the (unique!) relative velocity of a massive body
$q$ as measured/seen/ob\-ser\-ved by a massive laboratory/body $p.$
A category symbol $\varpi$ is interpreted as the physical process of
a measurement of a velocity. Two bodies in relative rigid rest are
considered as one object/body of groupoid category $\varpi.$

For any category in expression \eqref{display} one would expect to
write generally a membership $\uel\in\varpi(p,q)$ instead of an
equality $\uel=\varpi(p,q).$ However, in the groupoid category of
massive objects $\varpi,$ the cardinality of $\varpi(p,q),$ for
$\forall\,p,q\in\obj\varpi,$ by definition must be exactly one
$|\varpi(p,q)|=1.$ There is the \textit{unique} velocity of a
massive body $q$ relative to the reference system $p.$ The category
of observers $\varpi$ actually is an enriched category, a kind of a
$(1,1)$-category. Every velocity-morphism $\varpi(p,q)$ must be an
\textit{object} in a category $\V$ of all admissible relative
velocities, $\obj\V\equiv\{\varpi(p,q)\},$ and therefore $\varpi$
actually is a functor
\begin{gather}\begin{CD}\varpi\times\varpi @>{\varpi}>>\V.\end{CD}\end{gather}

Each object of $\varpi$ can be massive observer/source or
observed/target: there is no division between observer and observed.
However, in a velocity-morphism, $\varpi(p,q),$ a source object
$p\in\obj\varpi,$ actually is an observer/laboratory, and a target
object $q\in\obj\varpi,$ is a passive body being observed. Our
philosophy is that the relative velocity is objective and
independent of actual measurements.

In groupoid category of time-like vector fields (massive bodies) the
\textit{inverse} relative-velocity-morphism $\vel^{-1}$ is
interior-observer-dependent, and not absolute as it is in the
isometric-Lorentz-group formulation where, $\vel^{-1}\equiv-\vel.$
This is because the relative velocity $\vel$ is $g$-orthogonal to an
observer $p,$ $\vel\in\ker\,p,$ $p\vel=0.$ Somehow contrary to
orthogonality of relative velocity, Cui in very interesting paper is
emphasizing that rather the force $\simeq$ acceleration, and not
relative velocity, must be orthogonal to the particle-observer [Cui
2006], as it is in the electromagnetic theory with the accepted
Lorentz force-form $f=i_P F.$

The Einstein isometric ternary relative velocities, $\vel$ and
$\uel,$ see \eqref{ternary1}, are $\oplus$-add-able with respect to
the binary addition $\oplus,$ \ie $\exists\;\vel\oplus\uel,$ if and
only if these both velocities have the same preferred observing
laboratory $p$ as an implicit source as shown in Figure 1,
\begin{gather}\text{source}\,\uel\quad=\quad\text{source}\,\vel\quad=
\quad\text{source}\,(\vel\oplus\uel)\quad=p.\end{gather}

\begin{figure}[h]\caption{The nonassociative addition
$\oplus$ of isometric relative velocities needs preferred laboratory
$p.$}
$$\xymatrix{\bullet&\bullet\\
p\ar[u]^\vel\ar[ur]_{\vel\oplus\uel}\ar[r]_{\uel\oplus\vel}\ar[dr]_\uel
&\bullet\\\bullet&\bullet}$$\end{figure}
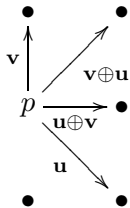

The Lorentz-boost-based non-associative addition $\oplus$ of
Einstein's isometric-velocities can be derived from the associative
addition of velocities-morphism given by the groupoid category of
observers $\varpi.$

It is believed that:\begin{itemize}\item The Lorentz isometry group
of transformations of orthogonal/Lorentz (1+1+1+1)-frames of
references is \textit{exclusive} for non-observable absolute
velocity.
\item The Lobachevsky coset space of the Lorentz group $SO(1,3)/SO(3)$
describe the space of relative velocities [Varicak 1910].
\item Lorentz-group-invariant physical laws is the \textit{highest}
guideline of physics. The concepts of Lorentz-covariance and the
Lorentz-invariance are discussed by Ivezic [2003] .
\item Nonassociativity of the addition
\eqref{Fock3} of isometric relative velocities is necessary for
physics, is unavoidable.
\end{itemize} We believe instead that:\begin{itemize}\item
The observer-independence $\not\equiv$ the Lorentz-group
relativistic invariance. The observer-dependence is not the same as
the Lorentz/Poincar\'e covariance. We think that Lorentz covariance
and invariance will dwindle in importance.
\item The change of the material reference system,
(1+3)-split/observer, need not to be representation of the isometry.
\item The isometric relative velocity (parameterizing Lorentz boost) needs
a preferred observer (it is a ternary link). Maybe physics do not
needs this?\item The binary relative velocities-morphisms are
unique, and therefore are added associatively.\end{itemize}

We are proposing the change of a concept of relative velocity, and
this gives an alternative explanation for nonassocia\-ti\-vi\-ty of
the addition $\oplus,$ and offer, among other, an alternative
solution for the Mocanu paradox in terms of the unique relative
velocities being the categorical morphisms.

The kinematics of the Lorentz-group-free relativity is ruled by an
operator algebra generated by atomic split-idempotents, and denoted
$\Obs(\varpi).$ We believe that this operator algebra is a Frobenius
algebra for each $n$-body problem, $\forall\;n\in\N,$ however the
full proof of this hypothesis is still missing.

A massive holonomic observer is a pair of the transverse equivalence
relations in the space-time:
\begin{itemize}\item Galileo Galilei (1564-1642) in his book in 1632
introduced the relation among events `to be in the same place'. This
relation defines the one-dimensional congruence and the codimension
one (three-dimensional) physical space of locations = positions as
the quotient space. To understand the relativity of the
`location=place' one must consider at least three massive events
from the point of view of two massive observers: an observer on the
beach, and a tourist on the traveling ship. One event alone
possesses neither a `place' nor a `moment' of a time!

Following Brillouin [1970] we emphasize the fundamental importance
of the non-zero-mass bodies for the formulation of the relativity of
the space and for the relativity of the (proper)-time. This is in
contradiction with wide spread opinion that the relativity theory
must be primary related to basis laws of massless optics and of
massless light propagation. That the Lorentz group is equivalent to
the massless light propagation, e.g. [Bargmann 1957]. Such opinion
originate not from Einstein formulation of special relativity in
1905, but from early Lorentz and Poincar\'e discovery in 1904 that
the Lorentz group is a symmetry of the Maxwell equations. In fact
the symmetry group of the Maxwell equations is \textit{not} Lorentz
group, and \textit{not} Poincar\'e group, but the conformal group
that does not found yet his proper place in the foundations of the
relativity theory. We agree that posteriori relativity theory can
and must be tested experimentally by means of the optical
experiments, however we claim the uselessness of massless radiation
for the definition of the Galilean relativity of space as the set of
locations. The massless radiation can not be considered to be the
reference system.
\item Albert Einstein in 1905 introduced the simultaneity relation among
events which define the relative proper time of the observer, and
the 1-co-dimensional quotient-congruence, and one dimensional the
Minkowski observer-dependent proper-time as the quotient [Minkowski
1908].\end{itemize}

These two equivalence relations must be seen as the pair of
transverse congruences, 1-dimensional and 1-co-dimensional: they
need the massive body as the primary concept. Massless radiation
does not possess the space-time split.

The Descartes and Newtonian motion, the relative velocity among
massive bodies, presuppose the concepts of (empty and absolute)
space (aether) and of time. If space is not absolute, then is the
massive observer-dependent. No matter $\Longrightarrow$ no concept
of the relative space. Therefore we wish to consider the concept of
the matter and the relative motion as the primary concepts that do
not need yet neither space nor time and no spacetime. Matter first.


\begin{thebibliography}{99}
\bibitem{Ive} Ivezi\'c Tomislav, The exact proof that Maxwell equations
with the 3D $\mathbf{E}$ and $\mathbf{B}$ are not Lorentz covariant
equations. The new Lorentz invariant field theory, physics/03011045.
\end{thebibliography}
\end{document}